\documentclass[conference]{IEEEtran}

\usepackage{floatflt}

\usepackage{color}
\usepackage{cite}

\usepackage[normalem]{ulem}
\usepackage{multirow}
\usepackage{graphicx}
\usepackage{amsmath}
\usepackage{amsthm}
\usepackage{amssymb}

\usepackage{verbatim}

\usepackage{psfrag,array}

\usepackage{amsmath}
\usepackage{epstopdf}
\usepackage{algorithmic}

\usepackage{booktabs}
\usepackage{footnote}
\usepackage{url}

\usepackage{xcolor}

\newcommand{\fo}{f_{\mathrm{o}}} % frequency offset 
\newcommand{\tof}{t_{\mathrm{o}}} % timing offset 
\newcommand{\Nf}{N_{\mathrm{f}}} % Number of grid points
\newcommand{\Ns}{N_{\mathrm{s}}} % 2400 samples in NPSS period
\newcommand{\Np}{N_{\mathrm{p}}} % 189 NPSS time domain sequence length

\newcommand{\p}[1]{\mathop{\mbox{\it p} } }

\renewcommand{\vec}[1]{\ensuremath{\boldsymbol{#1}}}
\newcommand{\be}{\begin{equation}}
\newcommand{\ee}{\end{equation}}
\newcommand{\ba}{\begin{array}}
\newcommand{\ea}{\end{array}}
\newcommand{\bea}{\begin{eqnarray}}
\newcommand{\eea}{\end{eqnarray}}
\newcommand{\bean}{\begin{eqnarray*}}
\newcommand{\eean}{\end{eqnarray*}}
\newcommand{\argmax}{\mathop{\arg\max}}

\newcommand{\rmt}{^{\rm T}}

\newcommand{\mr}{\mathrm}

\newcolumntype{L}[1]{>{\raggedright\let\newline\\\arraybackslash\hspace{0pt}}m{#1}}
\newcolumntype{C}[1]{>{\centering\let\newline\\\arraybackslash\hspace{0pt}}m{#1}}
\newcolumntype{R}[1]{>{\raggedleft\let\newline\\\arraybackslash\hspace{0pt}}m{#1}}

\definecolor{white}{rgb}{1,1,1}

\graphicspath{{./}}

\begin{document}

% \title{Hardware Accelerator for Fast and Energy-Efficient Timing Acquisition in NB-IoT Devices}
%\title{Hardware Accelerator for Maximum-Likelihood Timing Acquisition in NB-IoT}
\title{Maximum-Likelihood Detection for Energy-Efficient Timing Acquisition in NB-IoT}

\author
{
% \begin{tabular}{c}
% Harald Kr\"oll$^{\dagger}$,
% Matthias Korb$^{\dagger}$,
% Benjamin Weber$^{\dagger}$,
% Samuel Willi$^{\ddagger}$ and
% Qiuting Huang$^{\ddagger}$\\
% $^{\dagger}$Integrated Systems Laboratory, ETH Zurich, Switzerland \\
% $^{\ddagger}$Department of Electrical Engineering, University of Texas at Dallas, USA\\
% $^{\dagger}$\{kroell,mkorb,weberbe\}@iis.ee.ethz.ch, $^{\ddagger}$aldhahir@utdallas.edu\\
% \\
% \\
% \\
% \end{tabular}
\begin{tabular}{c}
Harald Kr\"oll$^{\dagger}$,
Matthias Korb$^{\dagger}$,
Benjamin Weber$^{\ddagger}$,
Samuel Willi$^{\ddagger}$ and
Qiuting Huang$^{\ddagger}$\\
$^{\dagger}$ACP AG, Zurich, Switzerland \\
$^{\ddagger}$Integrated Systems Laboratory, ETH Zurich, Switzerland \\
\{kroell,mkorb\}@newacp.ch
\{weberbe,huang\}@iis.ee.ethz.ch, 
\\
\end{tabular}
}

\maketitle

\begin{abstract}

Initial timing acquisition in narrow-band IoT (NB-IoT) devices is done by detecting a periodically transmitted known sequence.
The detection has to be done at lowest possible latency, because
the RF-transceiver, which dominates downlink power consumption of an NB-IoT modem, has to be turned on throughout this time.
Auto-correlation detectors show low computational complexity from a signal processing point of view at the price of a higher detection latency.
In contrast a maximum likelihood cross-correlation detector achieves low latency at a higher complexity as shown in this paper.
We present a hardware implementation of the maximum likelihood cross-correlation detection.
The detector achieves an average detection latency which is a factor of two below that of an auto-correlation method and is able to reduce the required energy per timing acquisition by up to 34\%.

% The low-latency cross-correlation method reduces the on time of the power-hungry RF-transceiver.

% Overall, the presented VLSI architecture of the timing acquisition unit increases the energy efficiency of NB-IoT cell search by a factor of three.

% We present a hardware accelerator, 
% %
% Considering the power constraint of an NB-IoT device we use an overlap-save cross correlation method which shows superior performance at the price of higher complexity.
% %
% The minimized detection time especially reduces the on time of the power-hungry RF-transceiver. Overall, the presented VLSI architecture of the timing acquisition unit increases the energy efficiency of NB-IoT cell search by a factor of three.

\end{abstract}

\section{Introduction}

Various estimates predict tens of billions devices connected to the Internet in 2020 in what is called the Internet of Things (IoT). IoT does not only take place in our homes or in areas which are covered by WiFi and other low-range networks, but also in remote places which are only covered by cellular or satellite networks. Cellular network coverage is almost ubiquitous and does not depend on proprietary end-user infrastructure. %Devices connected to the cellular IoT do not depend on IT infrastructure such as WiFi access points. Hence, they can be deployed more easily, since no changes in IT infrastructure are required and no third party equipment needs to be integrated. 

To realize an IoT in which the requirements for low-power, low-cost, and extended-coverage IoT devices will be met, the 3GPP consortium agreed on an LTE-Release-13 extension called Narrow Band (NB)-IoT or LTE Cat-NB1~\cite{nbiotprimer}.
On the downlink and uplink side NB-IoT mainly reuses LTE technology. However, cell search and timing acquisition procedures have undergone major adaptions to fit into the narrow 200\,kHz bandwidth and to
meet coverage extension requirements.

The energy efficiency of an NB-IoT device preferably implemented as a system-on-chip is of great importance to achieve years of battery life as aimed for emerging cellular IoT standards.
%~\cite{45820}.
Besides the power amplifier for the uplink, which holds the lions share of overall power consumption, it is well known that the downlink baseband signal processing consumes only a fraction of the RF-transceiver power in receive mode~\cite{ruf}.
This appears because RF-transceivers are dominated by analog integrated circuits whose power consumption especially does not scale as well with the CMOS technology feature size as it scales for the digital integrated baseband circuits.
Therefore, NB-IoT has undergone various simplifications to allow energy-efficient implementations.
%
% OLD
% Bandwidth reduction to 200\,kHz was the main simplification of NB-IoT compared to LTE. Besides that the maximum carrier frequency is 2.2 GHz and the adjacent channel leakage ratio was reduced by 5dB compared to 1.4MHz LTE~\cite{36101}.
% %
% Despite those simplifications, the RF-transceiver is still dominating the downlink power consumption because its power consumption is rather proportional to the carrier frequency and to sensitivity requirements than bandwidth.
% OLD END
% NEW
Significant bandwidth reduction to 200\,kHz was the main simplification of NB-IoT compared to the minimal bandwidth requirement of 1.4\,MHz in LTE. But, the RF-transceiver power consumption is rather proportional to the carrier frequency and to sensitivity requirements than bandwidth. While adjacent channel leakage ratio was reduced by 5dB compared to 1.4MHz LTE~\cite{36101}, the maximum carrier frequency is only slightly reduced from 2.6 to 2.2 GHz. Thus, the RF-transceiver is still dominating the downlink power consumption. 
%NEW END
%
However, power consumption of digital baseband processing scales well with bandwidth, which is useful for NB-IoT timing acquisition.

Besides data decoding timing acquisition is the most complex baseband
task along the downlink path~\cite{complexity_study}. Hereby energy-efficient 
timing acquisition is important because timing acquisition has to be done frequently, mainly for two reasons:
Firstly, NB-IoT is designed for the exchange of short messages, thus devices are in 
deep sleep mode most of the time and wake up e.g. every hour for a short period of time to receive and transmit 
a few hundred bytes. To ensure years of battery life, circuits providing accurate timing
are turned off during deep sleep mode, which requires timing acquisition after every wake-up.
Hereby timing acquisition has a relatively large share on the short reception interval, 
which requires an energy-efficient implementation.
Secondly, NB-IoT is likely to be used on vehicles and drones where devices are prone to timing synchronization loss due to their relatively high mobility and the absence of handover capability in NB-IoT.

%RF-transceivers in receive mode consume the major part of the power (up to 143\,mW in~\cite{borre} or around 203\,mW in~\cite{broadcom}). This also applies when receiving signals with a narrow bandwidth of 200\,kHz. 

For timing acquisition a periodically transmitted a priori known Narrowband Primary Synchronization Sequence (NPSS) has to be detected~\cite{36211}.
%Fast timing acquisition is of special importance for the energy efficiency of an NB-IoT device.
%
The latency of a successful timing acquisition (NPSS detection) is the relevant performance metric, because it determines how long the RF-transceiver, which consumes the major part of the power, has to be turned on to receive data. 
%The rest of the baseband (and protocol stack) is sleeping during detection, waiting for a \textit{wake-up signal} from NPSS detection. 
Therefore, using low-complexity NPSS detectors which achieve suboptimal performance can be disadvantageous for the overall downlink energy efficiency. 

% IoT like infrequent transmission and reception patterns of small amounts of data allows the the baseband to sleeping most of the time during detection, waiting for the NPSS detector to send a \textit{wake-up signal}

%MK Removed
%Furthermore, energy efficient timing acquisition is important because it is part of the cell-reselection procedure, which is the only mobility option available for NB-IoT devices which lack handover functionality.

\textit{Contributions:} We present a maximum-likelihood (ML) NPSS detector which achieves an average timing acquisition latency of 140\,ms (in-band deployment, TU1.2 channel, SNR = -12.6\,dB).
Our ML detector is based on cross-correlation metrics which are computed in frequency domain via the overlap-save method.
The detector has high computational complexity but allows to reduce the required energy by up to 34\% per timing acquisition for state-of-the-art RF-transceivers.
%
%
%The proposed VLSI implementation shows a diminishing power consumption compared to the RF-transceiver and thus increases the overall energy efficiency for NB-IoT timing acquisition.
%
% Outline
%We derive the overlap-save method for NB-IoT PSS detection, compare  complexity and performance against complexity and performance of a low-complexity detection method. In Section 5 we propose a VLSI architecture for the overlap-save method.

\section{Timing Acquisition in NB-IoT}

\begin{figure}
  \centering
  \includegraphics[width=\columnwidth]{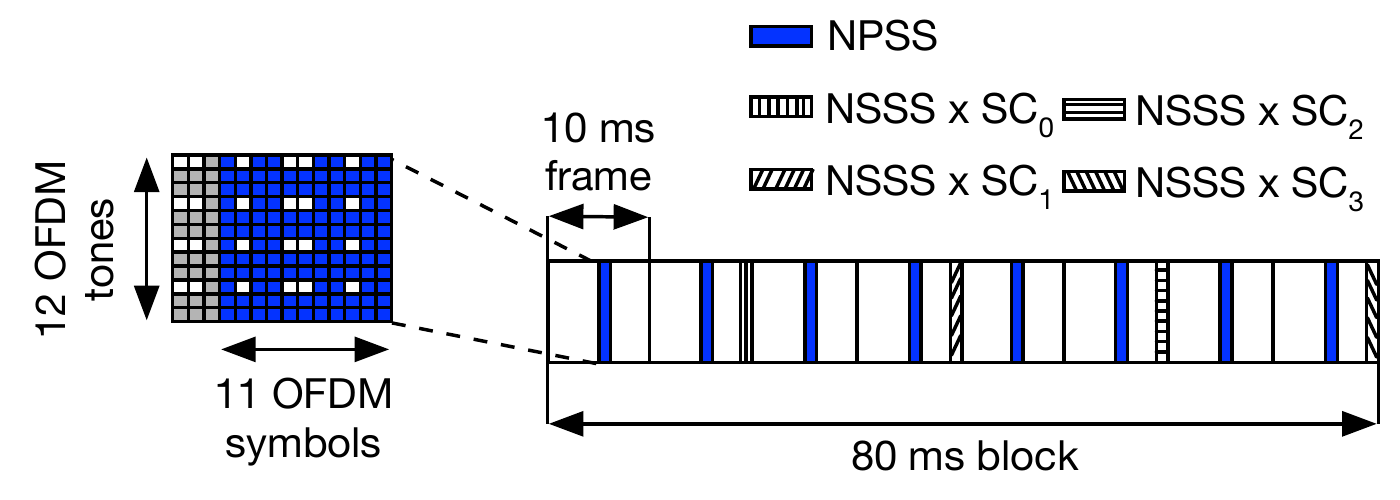}
  \caption{NPSS and NSSS resource mapping onto an NB-IoT frame.}
  \label{fig:resource_mapping}
\end{figure}

The first step after power-on (or after a wake-up from a sleep cycle) of an NB-IoT device is the detection of an NB-IoT capable base-station. In case such a base-station exists, the receiver does not know which OFDM symbol of the frame is currently transmitted. On top of that, the frequency relation between the base-station and the local receiver clock is also unknown. In NB-IoT as well as in other LTE device categories, the detection of a suitable base-station and the estimation of the timing and frequency offset is based on two periodically transmitted sequences: the NPSS and the Narrowband Secondary Synchronization Sequence (NSSS). While the NPSS is transmitted repeatedly every sub-frame of length 10\,ms, the NSSS is repeated in every second sub-frame as shown in Fig.\,\ref{fig:resource_mapping}. For NB-IoT the transmitted NPSS is identical in every sub-frame for all base-stations. In contrast, the NSSS depends on the base-station’s cell ID and is scrambled with a frame-dependent sequence code. 

%MK removed
%A base-station with a certain cell ID transmits four different SSSs in one 80\,ms frame.

The NPSS is used to verify the existence of an NB-IoT capable base-station. Additionally, it enables the estimation of the frequency offset and timing offset with respect to the sub-frame boundary. The NSSS is then used to detect the frame boundary and cell ID.

The NPSS is defined in frequency domain as a Zadoff-Chu sequence of length 11 for each sub-carrier index $n$ given by
\be 
S[n] = e^{ \frac{-j5\pi n(n+1)}{11} } c[l], \quad n = 0 \dots 10
\label{eq:pss}
\ee 
where $c[l]$ is an element of the code cover vector
\bean
\vec{c} = [1,1,1,1,-1,-1,1,1,1,-1,1],
\eean
with $l$ being the symbol index in a sub-frame.
This sequence is mapped to 11 subsequent OFDM symbols each consisting of 12 OFDM sub-carriers holding one copy of the NPSS. 

%MK removed
%The 11 copies slightly differ as some of the sub-carriers in certain OFDM symbols are reserved for pilots as indicated by white boxes in the tone allocation of Fig.~\ref{fig:resource_mapping}.
%

%
%Before the mapping onto the 12 sub-carriers of a certain OFDM symbol, the sequence is multiplied by a code cover S which can either take the value +1 or -1. 
After zero-padding each of the 11 copies to 128 symbols, time-domain conversion, and cyclic-prefix insertion of either length 9 or 10, the NPSS results in 1,508 time domain samples.

% \begin{figure}[!htb]
%     \centering
%     [Figure of PSS generation]
%     \caption{PSS generation.}
%     \label{fig:pss_generation}
% \end{figure}

With a sub-frame length of 10\,ms and a sampling rate of 1.92\,MHz 19,200 samples need to be captured in order to get exactly one copy of the NPSS. As the sub-frame boundary is unknown, the NPSS can start at any of the 19,200 samples. One task of the receiver is to estimate the beginning of the NPSS to acquire sub-frame boundary timing information. In addition, an NB-IoT device has a random frequency offset because the crystal oscillator on the device is not yet tuned after power-on or after wake-up from a sleep cycle. This heavily affects the detection complexity because the device needs to analyze various frequency-offset candidates within a specified boundary, as well. To reduce the complexity it is possible to perform a coarse frequency and timing offset estimation on a down-sampled version of the received signal. For example in~\cite{QC} the coarse estimation is done via auto-correlations at a sampling frequency of 240\,kHz. Then, one sub-frame consists of only 2,400 samples.

\begin{figure}[!t]
  \centering
  \includegraphics[width=1\columnwidth]{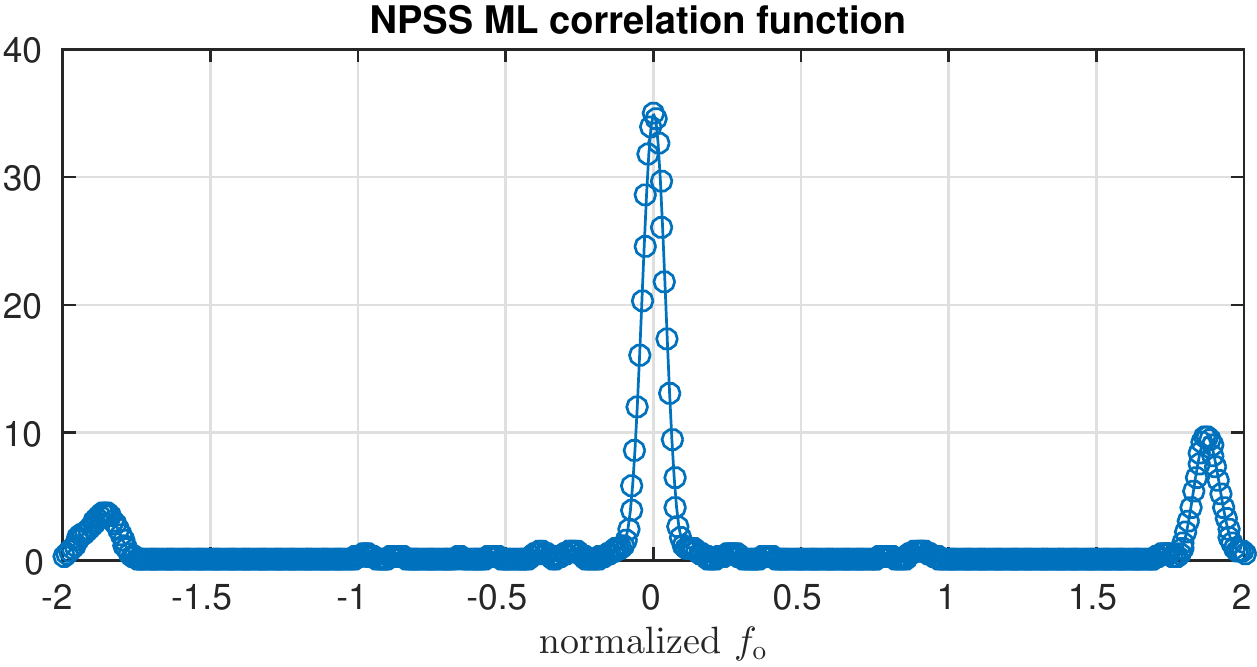}
  \caption{ML function over normalized frequency offset.}
  \label{fig:ccpeak}
\end{figure}

\section{ML Timing Acquisition with Correlations}

\begin{figure*}[!t]
  \centering
  \includegraphics[width=0.98\textwidth]{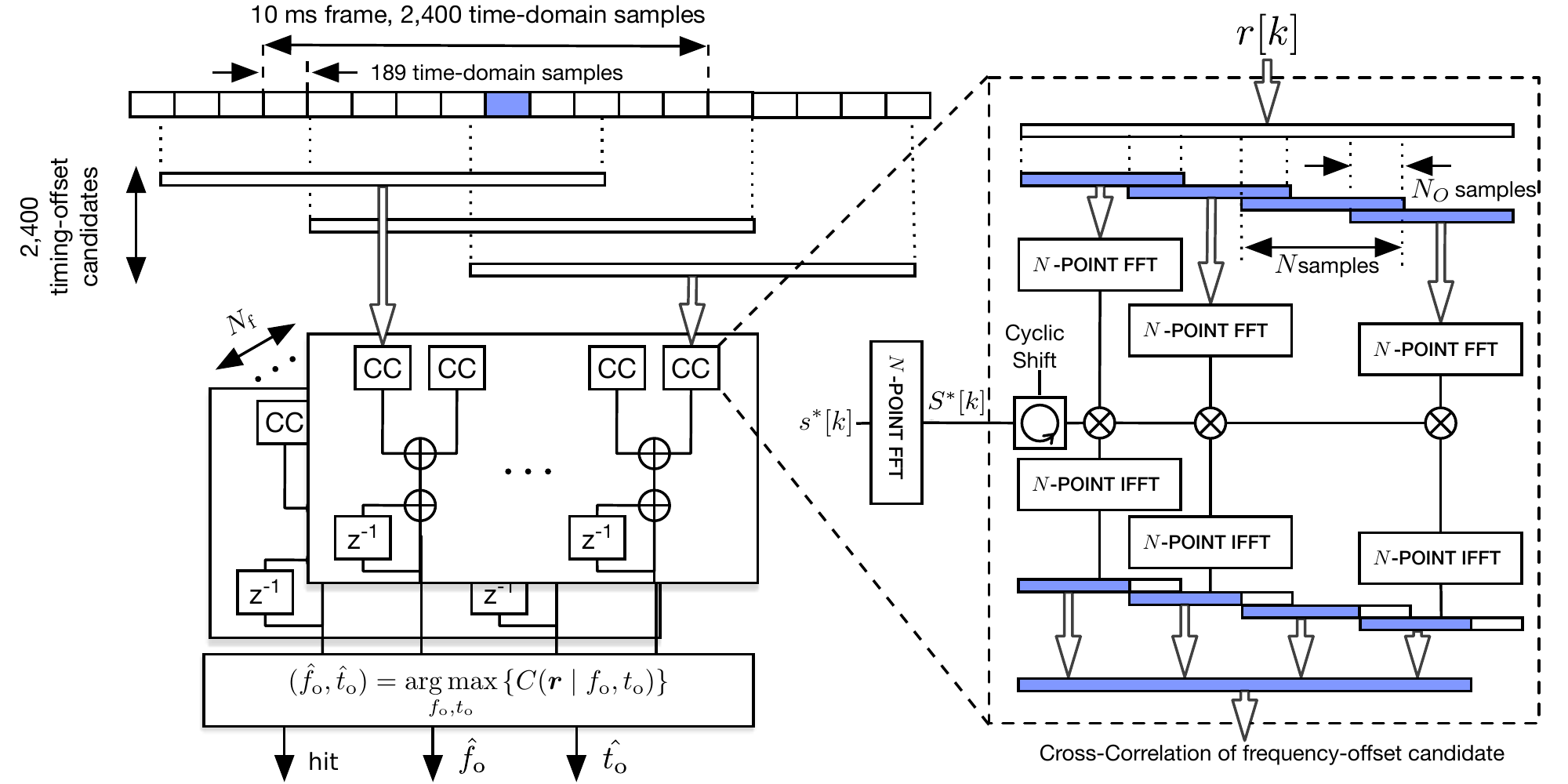}
  \caption{Signal processing scheme of the ML NPSS detector which includes fractional-frequency and coarse-timing offset estimation. The right sub-figure shows correlation computation with the overlap-save method. }
  \label{fig:cross_corr_algorithm}
\end{figure*}

There are two main algorithms to perform timing acquisition, namely auto-correlation and cross-correlation. While auto-correlation is the only option if the transmitted, periodic sequence is unknown, for NPSS detection both algorithms can be applied as the transmitted sequence is known to the receiver. Auto-correlation approaches are in general more hardware efficient than cross-correlation approaches. But, since the auto-correlation algorithm does not exploit the fact that the transmitted sequence is known, its performance is sub-optimal. In fact, cross-correlation detectors are ML detectors~\cite{proakis}. This is the reason, why many applications like radar systems or GPS receivers use a cross-correlation for signal detection~\cite{GPS}. 
%
% \subsection{High-Performance Low-Power Timing Acquisition}
In this paper we focus on low latency rather than low complexity. Thus, the ML detector~\cite{proakis} (Page 244), which projects the received signal vector onto each of the $\Nf$ possible frequency candidates, is a viable option for NPSS detection.

The NPSS ML detector correlation metrics are given by
\be
C\left( \vec{r} \mid \theta,f_{\mathrm{o}} \right) = \sum_{k=\theta}^{\theta+189} r[k+\theta]s^{*}[k] e^{-j 2\pi f_{\mathrm{o}}k/f_s},
\ee

where the received signal vector 
\bean
\vec{r}=\left[r[0] \; r[1] \;\ldots \;r[189-1]\right]\rmt    
\eean
has a sampling rate of $f_s =$ 240\,kHz and $s[k]$ for \mbox{$k = 0 \ldots 188$} is the time domain NPSS sequence given in Eq.~(\ref{eq:pss}) at 240\,kHz.

The ML function \mbox{$C( \vec{r}' \mid \theta=0,f_{\mathrm{o}} )$} for a distortion-free received signal vector $\vec{r}'$ over the frequency offset $\fo$ is plotted in Fig.~\ref{fig:ccpeak}.
The ML frequency- and timing-offset estimation $\hat{\fo}$ and $\hat{t}_{\mathrm{o}} $ can then be calculated according to
\bean 
( \hat{\fo},\hat{t}_{\mathrm{o}}  ) = \argmax_{\fo,\tof}{ \left\{ C(\vec{r} \mid \fo,\tof) \right\}. }
\eean

Hereby different time offset hypotheses $\tof$, which correspond to sub-frame boundaries, have to be evaluated by cross-correlating the received samples in the correlation window with the known NPSS.

In addition, correlations are performed for every $\Nf$ frequency offset
hypothesis $\fo$, which defines the range of frequency offsets the detector shall support.
%
% BW: removed as \Np gets introduced only in the subsequent sentence
%Hence, the number of cross-correlations in time-domain requires $\Np \cdot \Nf$ operations.
%
% TODO: later
The minimum frequency grid spacing is defined by the 240\,kHz sampling rate
and the FFT and IFFT size which trades off computational complexity, memory requirements, and processing delay for estimation accuracy.
Larger FFT sizes with smaller grid spacings improve frequency offset estimation accuracy but have a longer delay and require more memory.
An FFT size of 1,024 results in a grid spacing of 234\,Hz which
is sufficient for NPSS detection and was therefore chosen in this work.
%
% grid spacing
%
In addition, the width of the correlation peak of Fig.~\ref{fig:ccpeak} allows to take every fourth grid point only, while still covering 93\% of the peak amplitude.
$\Nf$ trades off the minimal observed height of a correlation peak against computational complexity and memory size.
It is a design parameter, which can be chosen to match the accuracy 
of the underlying crystal oscillator.
Choosing $\Nf=31$ leads to a frequency-offset range of $31 \cdot 4  \cdot 234\,Hz$ which allows to compensate $ \pm 14.5$\,kHz.

In every 10\,ms frame we receive \mbox{$\Ns$ = 2,400} samples and the length of the NPSS in time domain is \mbox{$\Np = 189$} samples. In total $\Ns \cdot \Nf$ cross-correlations of length $\Np$ are required
as shown in the left part of Fig.~\ref{fig:cross_corr_algorithm}. Considering \mbox{$\Nf=31$} different frequency candidates a total of 74,400 cross-correlations need to be performed every 10\,ms, which is impractical for NB-IoT devices.

However, the computational complexity can be significantly reduced when using an overlap-save (OLS) method~\cite{overlapsave}. This method is well established especially for discrete convolutions but it can also be applied to cross-correlations. 
By applying OLS to the NPSS cross-correlation, the input stream is divided into overlapping sequences of length $N$ as illustrated in the right part of Fig.~\ref{fig:cross_corr_algorithm}.
The number of overlapping samples depends on the NPSS length and is chosen to be $N_O=188$. Afterwards, the block-wise cross-correlation with the different frequency-offset candidates is performed, which can be done in frequency domain. The main benefit of this method is that a cross-correlation in time domain is replaced by a point-wise multiplication in frequency domain. 
Additionally, the generation of the NPSS reference signals in frequency domain gets simplified: Different frequency offsets relate to cyclic shifts which can be easily implemented in hardware.

%$\Nf = 31$ still leads to an adequate correlation on the neighboring frequency boundaries, even if the frequency offset lies exactly between the two hypothesis. 

%A fast detection of the  can significantly reduce the receiver power consumption especially because the baseband processing only consumes a fraction of the total receiver power as indicated earlier. From this perspective, replacing the auto-correlation in the low-complexity timing acquisition algorithm by a cross-correlation is promising. Note that the second and third step of the low-complexity timing acquisition based on cross-correlations remain unchanged. 

\section{Complexity and Performance}

%In this section we analyze and compare the complexity in terms of MOPS of the low-complexity timing acquisition~\cite{QC} and the proposed OLS cross-correlation method. For the low-complexity method the computational complexity and memory requirement is dominated by the auto-correlation which represents the first step of the proposed timing acquisition. The complexity for this auto-correlation is reported to be 25.6 MOPS and the memory requirement as 24 KB.

% We need 1 FFT and Nf IFFT per 2400 samples
% Each FFT has N/2*log2(N) *4 multiplications and as many additions
% The multiplication in freq domain requires 4 * N real multiplications and 2 * N real additions

\begin{figure}[!t]
  \centering
  \includegraphics[width=\columnwidth]{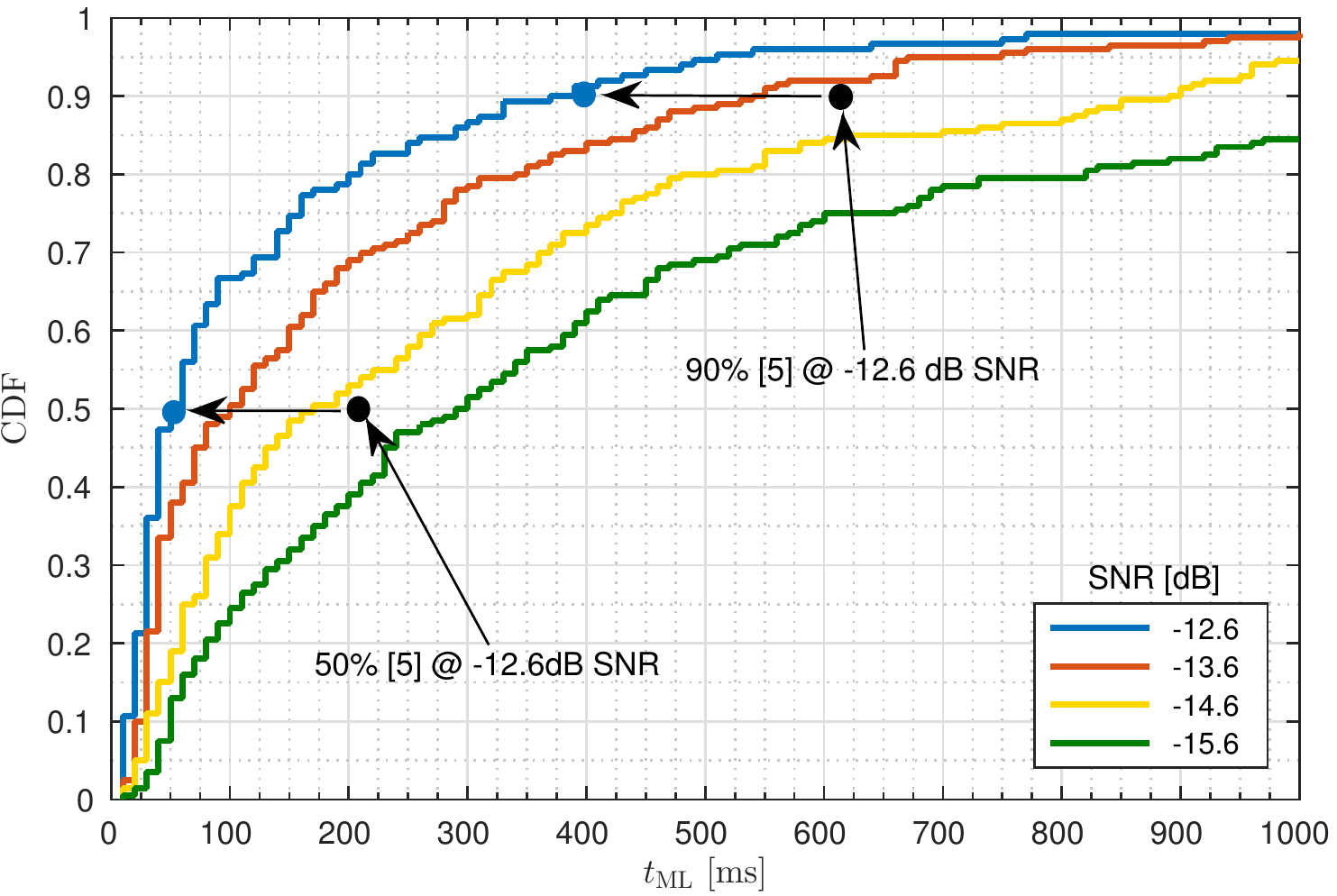}
  \caption{Timing acquisition latency of cross-correlation NPSS detector.}
  \label{fig:CDF}
\end{figure}

As the proposed cross-correlation-based algorithm is an ML detector, the 
complexity is expected to be significantly higher compared to the low-complexity auto-correlation method. On average for each received block of size $N-N_O$ a single $N$-point FFT, $N_f$ point-wise complex multiplications of a vector of length $N$, and $N_f$ $N$-point IFFT operations need to be performed.  Choosing an FFT size of $N=1,024$ the number of real additions and multiplications can be estimated to 135.0 and 135.4~MOPS, respectively leading to an overall computational complexity of 270.5 MOPS. Thus, the computational effort per sub-frame of the ML detector is roughly 10x higher than the auto-correlation timing acquisition~\cite{QC}.

%[TODO: Add numbers with MOP/detection here and in table 2]

%MK removed
%As we will see in the following, the ML detector does not need as many sub-frames as the low-complexity detector. Hence, its overall computational effort per timing acquisition lies actually below 10x that of the low-complexity timing acquisition.

% i don't get this:
% Another challenge for NB-IoT timing acquisition is the very low sensitivity requirements of (???Value, Metric). At these noise levels it is not sufficient to do one auto-correlation within one sub-frame. It is rather imperative to increase the auto-correlation window to multiple sub-frames. 

The performance in terms of timing-acquisition latency is shown in Fig.~\ref{fig:CDF} for in-band 
%\myworries{BW: in-band is mentioned here for the first time, remove it and just mention the SNR}
deployment which has the most demanding SNR requirement of -12.6\,dB and beyond.
For the simulations the TU1.2 channel model was used and the threshold was set to achieve a false-alarm rate of 1\%. 
The OLS detector
achieves a latency of 400\,ms, whereas the auto-correlation detector
of~\cite{QC} takes 620\,ms to achieve a 90\% hit rate.
The average detection latency is 140\,ms which is roughly a factor of two below
the value of~\cite{QC}.

%MK removed
%The increased computational complexity can be addressed by a VLSI implementation which leads to a lower energy consumption as shown in the following.

\section{Hardware Implementation}

\begin{figure}[!b]
  \centering
  \includegraphics[width=\columnwidth]{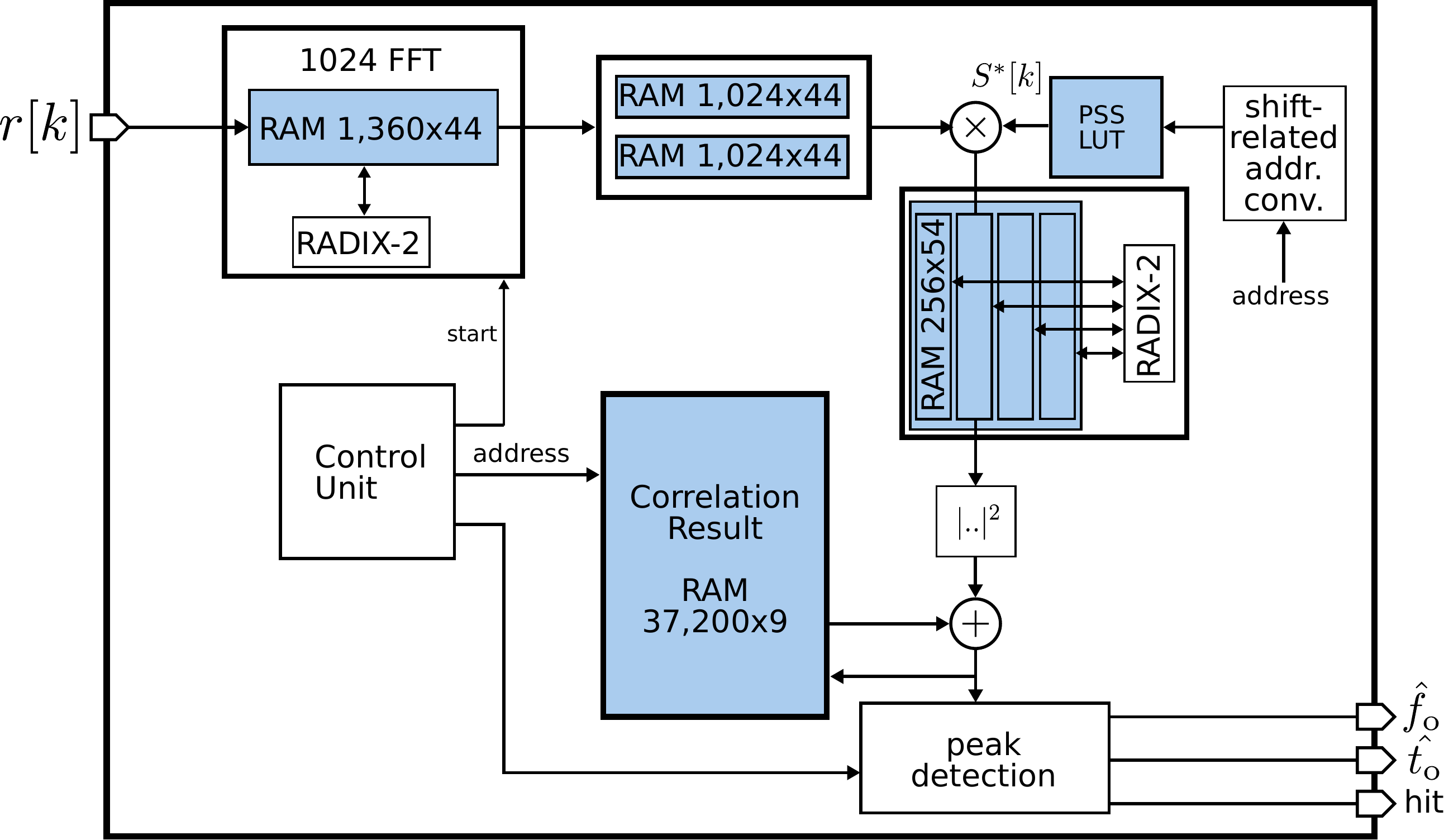}
  \caption{Block diagram of NPSS detector.}
  \label{fig:blkdiag}
\end{figure}

A block diagram of the cross-correlation NPSS detector is shown in Fig.~\ref{fig:blkdiag}. The main computational elements are the FFT and IFFT blocks with a required throughput of $2.87/10\,\mathrm{ms}=287.1/\mathrm{s}$ and $890.0/\mathrm{s}$ FFT and IFFT computations or $1.5$ and $45.6$ million radix-2 operations per second, respectively. Even for the more demanding IFFT it is possible to reuse a single radix-2 instance for all IFFT operations when assuming typical VLSI clock frequencies. So, for the FFT as well as for the IFFT block a single radix-2 in-place architecture is sufficient. 

The FFT is designed to include a RAM holding 1,360 complex samples which is larger than $N$. The reason for this is two-fold: Firstly, the FFT operates on 1,024 complex words, but the unaltered 188 overlap samples need to be stored for the next FFT computation, as well. Secondly, during the FFT operation further inputs $r[k]$ need to be stored in the memory. Furthermore, a single-port RAM has been chosen, which minimizes the storage area. The introduced memory-bandwidth bottleneck limiting the throughput to one radix-2 operation every 4 clock cycles is tolerable due to the very low throughput requirements of the FFT.

In contrast such an architecture would not be sufficient to meet the throughput requirement of the IFFT. Here, the memory bandwidth has to be 4$\times$ higher to support a throughput of one radix-2 operation every cycle. Thus, the memory in the IFFT block is split into four banks each still being a single-port RAM to minimize storage area. Memory access conflicts are avoided by assuring that every two subsequent radix-2 operations do not access the same register banks. 
After processing the FFT, the correlations in frequency domain, and the $N_f=31$ IFFTs for each received block of length $N$ the results are non-coherently combined with previous correlation results. The size of the memory holding the intermediate, non-coherently combined correlation results is reduced by down-sampling the correlation results by a factor of 2 as proposed in~\cite{QC}. After the processing of a sub-frame, a peak-detection is used to decide, whether the NPSS sequence was found. Rather than using a simple peak-to-average ratio an analysis of the four largest correlation results is considered which improves the detection probability when combining correlation results of multiple sub-frames. Also, the existence of side-peaks (Fig.~\ref{fig:ccpeak}) requires a more sophisticated peak detection as a simple peak-to-average ratio would lead to many false detections.

%
% Optional sentence:
%One result ready every 10ms, which is reported to the attached processor via an interrupt signal.

We implemented the detector in VHDL and performed synthesis experiments in SMIC130 and GF28 CMOS technology targeting a clock frequency of 62\,MHz.
The key characteristics of the detector are give in Table\,\ref{tbl:char}.

\begin{table}[!t]
\small
  \begin{center}
    \renewcommand{\arraystretch}{1.2}
    \caption{NPSS Detector characteristics in two CMOS technologies.}
    \label{tbl:char}
    %\begin{tabular}{|l||c|}
    \begin{tabular}{@{} c c @{} >{\kern\tabcolsep}c @{}}
      \toprule
    CMOS technology   & SMIC 130\,nm        &  GF 28\,nm\\ 
      \midrule
    Synthesized Cell Area              & 3.34\,mm$^2$      & 0.22\,mm$^2$     \\ 
    Voltage           & 1.2\,V              & 1.0\,V     \\ 
    kGE               & 735          & 600     \\ 
    est. $P_{\mr{ML}}$     & 38\,mW       & 2.5\,mW    \\ 
      \bottomrule
    \end{tabular}
  \end{center}
\end{table}

With $N_f=31$ a correlation RAM with 334\,kbit is required.
This is the largest memory in the design and occupies
54\% of the entire area.
However since this memory is only used for NPSS detection it can be
easily shared with other building blocks.
The implementation also includes the fine frequency- and timing-offset estimation as proposed in \cite{QC}.

The power consumption of the detector was estimated by using Cadence\textsuperscript{\textregistered} tools from post-synthesis netlist and
the value change dump file to 38\,mW (1.2V, TT, 25C) and 2.5\,mW (1.0V, TT, 25C) for the 130- and 28-nm technology, respectively.

\section{Energy Efficiency}

The energy of timing acquisition is given by the power of the detector
and the RF-transceiver in receive mode times the latency $t$. Given
the energy of the ML approach and the auto-correlation (AC) approach for a certain RF-transceiver power $P_{\mr{RF}}$ we compute the savings according to
\begin{equation*}
\Delta E \,[\%] = 100
\left[
1 - 
\frac
{(P_{\mr{RF}} + P_{\mr{ML}})t_{\mr{ML}}}
{(P_{\mr{RF}} + P_{\mr{AC}})t_{\mr{AC}}}
\right].
\end{equation*}
For the AC timing acquisition we account for a power of $P_{\mr{AC}}=\frac{P_{\mr{ML}}}{10}$ because the arithmetic load is about 10$\times$ below the arithmetic load of the ML approach. However it shall be denoted that this factor is dependent on the implementation.

\begin{figure}[!ht]
  \centering
  \includegraphics[width=\columnwidth]{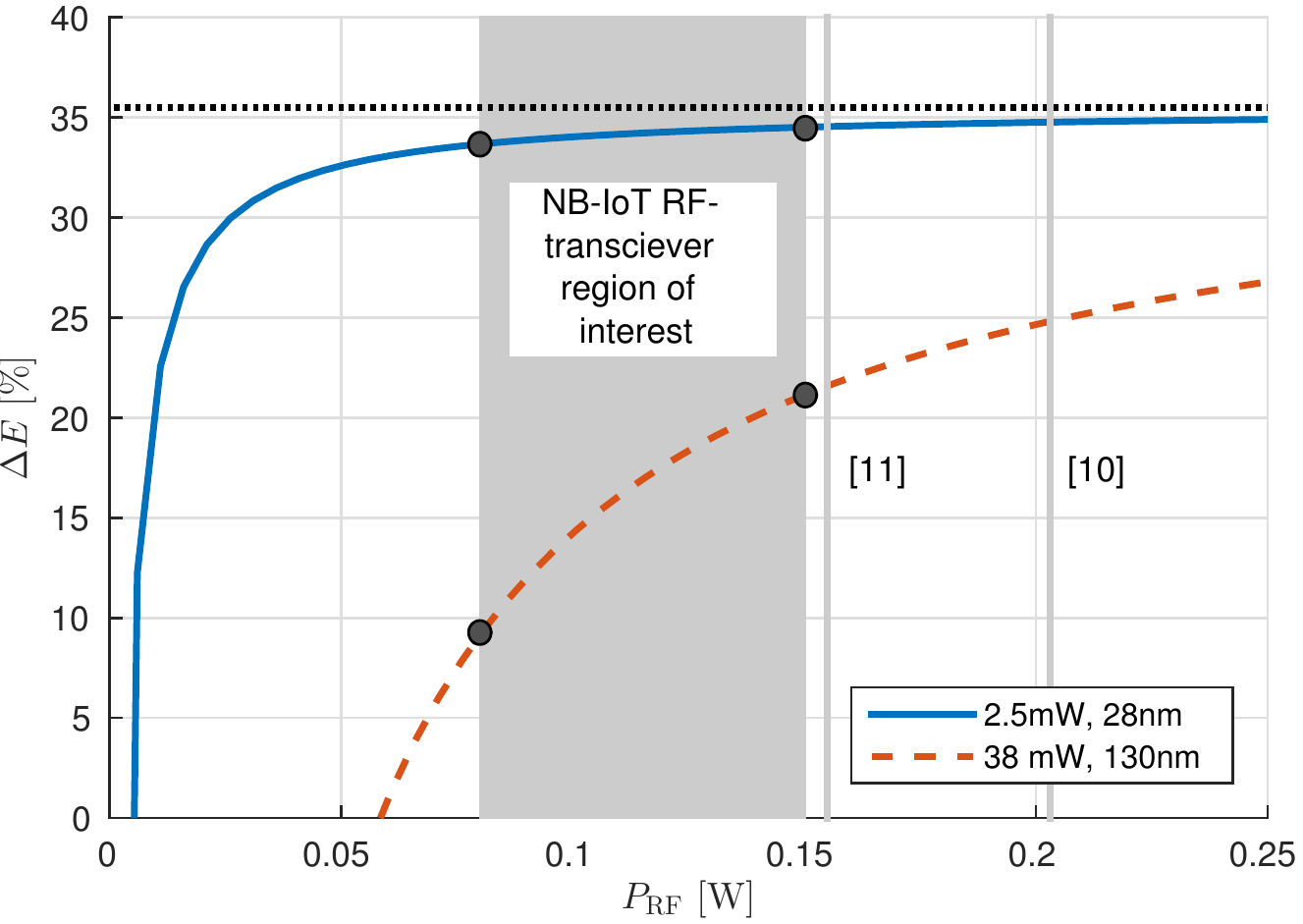}
  \caption{Energy savings per timing acquisition for the ML detector with
  400\,ms latency over the auto-correlation detector 620\,ms latency with different power consumption values.}
  \label{fig:E}
\end{figure}

In Fig.\,\ref{fig:E} the energy saving $\Delta E \,[\%]$ per timing acquisition is plotted over the power consumption of the RF-transceiver $P_{\mr{RF}}$ [W] for the latency of the ML detector ($t_{\mr{ML}} = 400$\,ms) and the AC detector ($t_{\mr{AC}} = 620$\,ms) in~\cite{QC}.

Even though the AC detectors show a lower power consumption 
for NPSS detection (due to their reduced number of additions and multiplications) they do not improve overall energy efficiency
because of higher latency.
The dotted line shows the maximum possible savings of 35.5\%.

The power consumption of RF-transceivers is dependent on multiple factors whose analysis lie beyond the scope of this paper, therefore we
consider a broad range of values for RF-transceiver power consumption.
The grey rectangle in Fig.\,\ref{fig:E} indicates the region of
interest for NB-IoT dedicated RF-transceivers which lies below
the power consumption of conventional LTE and GSM transceivers
due to the simplifications made in NB-IoT.
Power consumptions of state of the art conventional LTE and GSM transceivers are indicated by the vertical lines in Fig.\,\ref{fig:E} indicate the power consumption of two reported RF-transceivers~\cite{ericsson,broadcom}.

%Furthermore, it can be seen from Fig.\,\ref{fig:E}, that a low-latency NPSS detection leads to energy savings for RF power consumption values above 60\,mW.

%

%Naturally, the power consumption of the auto-correlation method depends on its implementation. To make a fair comparison, we assume that the low-complexity auto-correlation can be implemented efficiently in VLSI and thus has a very low power consumption. \myworries{BW: the table suggests an auto-correlation power consumption of 1/10 of our approach, a few sentences which justify this power consumption by means of the 10x lower computational complexity is needed.}

%\myworries{BW: this section lacks a bold statement reiterating the energy/detection from the table, what's more, the table is currently not even mentioned}

% \section{Note}
% Frequency drift of the oscillator is not considered in our detector
% because during the maximum acquisition time (x ms) the oscillator frequency
% is assumed to be constant. This is actually OK for temperature compensated oscillators (TCXO).

% It shall be denoted that only the dedicated VLSI implementation brings the presented efficiency. If the OLS method would be implemented on a DSP which has to compute 704\,MOPS/s the savings would be mitigated and the low-complexity auto-correlation method would be more feasible.

\vspace{0.5cm}

\section{Conclusion}

The fact that the RF-transceiver dominates downlink power consumption
in an NB-IoT device creates design space for dedicated hardware implementations which can execute exhaustive baseband algorithms.
Following this guideline we have shown that the computationally complex ML approach for NB-IoT timing acquisition can lead to significant energy savings in NB-IoT devices.
The savings were achieved by the low latency of our detector which due to algorithmic transforms based on the OLS method and by targeting a dedicated VLSI implementation shows a relatively low power consumption.
We were able to reduce the energy required for a single NPSS detection by 34\% for 28\,nm CMOS technology and from 9\% up to 21\% even in a rather mature 130\,nm CMOS technology.
Future research will address area reductions especially by sharing memory resources with other hardware building blocks.
%
%\myworries{BW: I thought we intend to compare an auto-correlation VLSI with our cross-correlation VLSI (at least this is what the last paragraph of the previous section suggests), meaning that the second statement of the previous sentence kinda looses it significance. the next sentence brings it to the point, why we actually save: less duration, less Rx-on!}
%

\bibliographystyle{IEEEtran}
%\bibliography{bibliography}

\end{document}